\documentclass[twocolumn,showkeys,showpacs,preprintnumbers,amsmath,amssymb]{revtex4}
\usepackage{amsmath}
\usepackage{amssymb}
\usepackage{epsfig}

\begin{document}

\def\bea{\begin{eqnarray}}
\def\eea{\end{eqnarray}}
\def\bra#1{\mathinner{\langle{#1}|}}
\def\ket#1{\mathinner{|{#1}\rangle}}
\newcommand{\braket}[2]{\langle #1|#2\rangle}
\def\Bra#1{\left<#1\right|}
\def\Ket#1{\left|#1\right>}

\floatsep 30pt
\intextsep 46pt

\title{RELATIVISTIC PSEUDO GAUSSIAN OSCILLATORS}
\author{Felix Iacob}
\email{felix@physics.uvt.ro, felix.iacob@gmail.com} 
\affiliation{{\it West University of Timi\c soara V P\^ arvan Ave 4,\\
RO 300223 Timi\c soara Romania, \\felix@physics.uvt.ro}}

\begin{abstract}

The quantum models of a massive scalar particle inside of an open bag
generated by a pseudo-Gaussian conformaly flat (1+1) metrics are investigated.
The potential of a free moving test particle, in the generated metric, has Gaussian 
asymptotic behavior, approaching to the potential of harmonic oscillator in the limit of zero.
The energy levels are calculated using numerical methods, calculations are based on
efficient method of generating functionals.
\end{abstract}

\keywords{Hamiltonian system, Gaussian potential, Generating functionals, Energy levels.}

\pacs{ 04.62.+v, 03.65.Ge }

\maketitle 

The non-relativistic harmonic oscillator (HO) is one of the simplest and most
useful system in physics. However, its relativistic generalization
is not well defined. First attempt were given by Yukawa\cite{Yukawa}
and further developed by several authors\cite{oscil}. These attempts
were based on covariant generalization $x_\mu x^\mu$ of non
relativistic potential. The oscillator based on Dirac equation were
also analysed  by Moshinsky\cite{Mdo} et al. Other proposals were
based on the generalization of the symmetry algebra of quantum
operators of quantum free systems \cite{NavarroA}. Furthermore,
geometric models with different metrics were used to simulate
relativistic oscillators \cite{Nav}, \cite{CotRHO}. Systems with
harmonic oscillator \cite{CGaussBag} and relativistic harmonic
oscillator \cite{RGaussBag} behaviour can also be induced by a
Gaussian potential. Recently  a new proposal \cite{PGO} was made by
introducing pseudo-Gaussian potentials to describe a new family of
quantum models with HO behavior when approaching to zero. Here we
would like to investigate the relativistic case of this recently
proposed models, by using pseudo-Gaussian potentials. We have to
remark that all models \cite{CotRHO}\cite{RGaussBag}\cite{PGO} have
the same limit: the nonrelativistic HO. This indicates that this
work could be done in the energy basis of HO. This means that the
pseudo-Gaussian potential must behave as HO one in some cases. We
will see that with suited coefficients this will be accomplished.

The matrix elements of Hamiltonian operator can be calculated in the
energy basis of HO. This can be done by using the method of
generating functions \cite{Ball}.

Let's consider the quantum model of a scalar test particle with the
mass $M$ freely moving inside a pseudo-Gaussian bag, simulated by
the (1+1) geometry, defined in a manifold within a local static
chart by the line element:

$$
ds^2 = g_{00}dt^2+g_{11}dx^2=g(x)(dt^2-dx^2)
$$
where
\begin{equation}
g(x)=1+v(x)=1+\left ( \lambda + \sum^r_{i=1}C_i x^{2i} \right
)exp(-\omega^2x^2)\label{metr}
\end{equation}
with $\omega>0$ and $\lambda$ an arbitrary ground energy parameter,
used to establish the depth \cite{PGO} of the potential well .
Taking into consideration some further conditions on the
coefficients $C_i$, they will be introduced latter.  Considering an
observer in $x=0$ and since the metric is static the energy is
conserved, so the Klein-Gordon equation for the free test particle
writes $[\square+M^2]\Phi =0$ and allows solutions of the form:

\begin{equation}
\Phi^{(+)}(x,t) = \frac{1}{\sqrt{2E}}U_E(x)e^{-iEt}; \,\,
\Phi^{(-)}=(\Phi^{(+)})^*.
\end{equation}

Thus  it is obtained:

\begin{equation}
\left [ -\frac{d^2}{dx^2} + V(x) \right ]U_E(x)=(E^2-M^2)U_E(x)
\label{kgx}
\end{equation}
where the potential \cite{PGO} is:
\begin{equation}
V(x)=  M^2 \left ( \lambda + \sum^r_{i=1}C_i x^{2i} \right
)exp(-\omega^2x^2) \label{pot}
\end{equation}

The free motion in the metric (\ref{metr}) can be seen as one in the
potential (\ref{pot}). Thus according to equation (\ref{kgx}) the
energy spectrum has a discrete part in the domain $[0,M)$ and a
continuous one for $E>M$. The eigenfunctions of discrete part must
be square integrable,

$$
\braket{U_{E^\prime}}{ U_E}=\int dx[U_E(x)]^*U_{E^\prime}(x)
$$

The problem of finding the discrete energy levels cannot be solved
explicitly in the potential (\ref{pot}). However, we can
identify the energy levels step by step up to an arbitrary order,
using computational methods. In this way, introducing the variable
$\xi=\sqrt{M\omega}x$ and the parameters $k$ respective $\nu$ by:

\begin{equation}
k=\frac{M}{\omega},\; \;\; 2\nu
+\lambda(k-1)+1=\frac{E^2-M^2}{M\omega} \label{ev}
\end{equation}
the equation (\ref{kgx}) becomes:
\begin{equation}
\Delta^r_{\lambda \, k} U_\nu = (2\nu+1)U_\nu\label{evp}
\end{equation}
where $U_\nu$ denotes $U_E$. The operator from the left side of the
above equation is:

\begin{equation}
\Delta^r_{\lambda \, k} =-\frac{d^2}{dx^2} + W^r_{\lambda \,
k}(\xi)-\lambda(k-1) \label{op}
\end{equation}
and the potential (\ref{pot}) become:

\begin{equation}
W^r_{\lambda \,k}(\xi)=k\left ( \lambda + \sum^r_{i=1}C_i \xi^{2i}
\right )exp(-\frac{\xi^2}{k})-\lambda(k-1) \label{potxi}
\end{equation}
and it is supposed  to exhibit Taylor expansions,  as follows:

\begin{equation}
W^r_{\lambda \, k}(\xi)=\xi^2+ {\cal O}(\xi^{2r+2})
\end{equation}
without terms proportional to $\xi^4,\xi^6, \ldots, \xi^{2r} $, thus
the potential have a Gaussian asymptotic behavior and in the
neighborhood of $\xi=0$ will behave like HO potential.
Furthermore, in the case of nonrelativistic limit, the potential
(\ref{potxi}) have to be such that the operator (\ref{op}) becomes
the HO operator $ \Delta_\lambda= -\frac{d^2}{dx^2} +\xi^2+\lambda$.
These two conditions are accomplished if the coefficients $C_i$ have
the following form:

\begin{equation}
C_i:=\frac{\lambda+i}{k^i i!} \label{coef}
\end{equation}

In consequence it is easy to verify that:

\begin{equation}
\lim_{r\to +\infty} W^r_{\lambda \, k}(\xi) =\xi^2+\lambda \;
\Rightarrow \; \lim_{r\to +\infty} \Delta^r_{\lambda \,
k}=-\frac{d^2}{dx^2} +\xi^2+\lambda \label{limr}
\end{equation}
and

\begin{equation}
\lim_{k\to +\infty} \Delta^r_{\lambda \, k}=-\frac{d^2}{dx^2}
+\xi^2+\lambda. \label{limk}
\end{equation}

The potential (\ref{potxi}) together with coefficients (\ref{coef})
is named pseudo-Gaussian potential, changing $r$ and $\lambda$ a
class  of pseudo-Gaussian potentials, denoted $(k,\lambda)^r$, are
obtained \cite{PGO}. We have to point out that the coefficients
(\ref{coef}) are not uniquely defined, however the coefficients used
in \cite{PGO} do not satisfy condition (\ref{limk}). Taking into
consideration the behavior (\ref{limr}) and (\ref{limk}) of the
potential it is convenient to work in the energy basis of HO,
$\{\ket{n} | n=0,1,\ldots\}$ In this basis the operator
$\Delta^r_{\lambda \, k}$ could be put in diagonal form in the
desired approximation by fixing $n=N$. A ($N\times N$) truncated
matrix is obtained, thus we can solve the eigenvalue problem
(\ref{evp}) by calculating the numeric values of $\nu$ which
correspond to the discrete energy levels $\epsilon_n=2\nu_n +1$. In
this parametrization $\nu$ is similar to the quantum number of HO
and its values verify the condition:

\begin{equation}
2\nu +1<-\lambda(k-1)\label{nucond}
\end{equation}
The values for $\nu$ are close to integers and will see that the
particle behaves like a HO one. To find this values we will use
perturbations theory, in the energy basis of HO, using the technique
of generating functions.\\

 The generating function, in this case can be written as follows:

\begin{equation}
F_\tau (\xi)=\left ( \frac{1}{\pi}\right )^{1/4}  exp\left (
-\frac{\xi^2}{2}+ 2\xi\tau+\tau^2\right )
\end{equation}
which yields the HO eigenfunctions normalized in the $\xi$-scale as:

\begin{equation}
u_n(\xi)=\braket{\xi}{n}=\frac{1}{\sqrt{(n!2^n)}} \left .
\frac{d^{n}  F_\tau (\xi)}{d\tau ^{n}}\right |_{\tau=0}
\end{equation}

The matrix elements of any operator ${\cal X}$, in this basis, can
be derived from the related generating functional,

\begin{equation}
Z_{\sigma,\tau}[{\cal X}]=\int d\xi F_\sigma (\xi)[{\cal
X}F_\tau](\xi) \label{fg}
\end{equation}
according with:

\begin{equation}
\bra{n}{\cal X}\ket{m} = \left . \frac{1}{\sqrt{n!m!2^{n+m}}}\,
\partial_\sigma^{n}\partial_\tau^{m}Z_{\sigma,\tau}[{\cal X}]\right
|_{\sigma=\tau=0} \label{elem}
\end{equation}

In general the integral (\ref{fg}) reduces to some known Gaussian
integrals. For example in the case of HO:
$Z_{\sigma,\tau}[\Delta_\lambda]=(1+\lambda+4\sigma,\tau)exp(2\sigma,\tau)$
with the matrix elements $\bra{n}{\cal X}\ket{m}
=(2n+\lambda+1)\delta_{m\,n}$.

The generating functional of an arbitrary model $(\lambda,k)^r$ is
given by:

\begin{equation}
Z_{\sigma,\tau} [\Delta^r_{\lambda \, k}]= Z_{\sigma,\tau}[
-\tfrac{d^2}{d\xi^2}]+Z_{\sigma,\tau}[W^r_{\lambda,k}-\lambda(k-1)]
\end{equation}
the terms that can be calculated as Gaussian integrals are:

\begin{equation}
Z_{\sigma,\tau}[ -\tfrac{d^2}{d\xi^2}]=
[\tfrac{1}{2}-(\sigma-\tau)^2]\,exp(2\sigma\tau),
\end{equation}

\begin{equation}
Z_{\sigma,\tau}[exp(-\tfrac{\xi^2}{k})]=\frac{\sqrt{k}}{\sqrt{k+1}}\,
exp \left ( 2\sigma\tau -\frac{1}{\sqrt{k+1}}(\sigma+\tau)^2\right),
\end{equation}
using the identity $ \xi^{2i} e^{-\tfrac{\xi^2}{k}} =
[k^{2i}\partial^i_k ] e^{-\tfrac{\xi^2}{k}}$ to rewrite the
generating functional for the potential term, we can conclude:

\begin{eqnarray}
Z_{\sigma,\tau} [\Delta^r_{\lambda \, k}] &=& \left (
\frac{1}{2}-(\sigma-\tau)^2 -\lambda(k-1)\right )\,exp(2\sigma\tau)
\label{ourgf}
\\\nonumber & +& k\left ( \lambda + \sum^r_{i=1}C_i [k^{2i}\partial^i_k ]
\right ) \\\nonumber &\times&
\left \{ \frac{\sqrt{k}}{\sqrt{k+1}}\, exp \left (
2\sigma\tau -\frac{1}{\sqrt{k+1}}(\sigma+\tau)^2\right ) \right
\}
\end{eqnarray}

The expression (\ref{ourgf}) for the generating functional will be
used to compute the matrix elements of the operator
$\Delta^r_{\lambda \, k}$. The methodology used will be described
below as follows:

A concrete value for $r$ have to be taken. The value for $r$
determine the shape of the potential (\ref{potxi}). A value $r=3$
have been considered. By fixing $\lambda$, we have a concrete model
$(k,\lambda)^r$ i.e. the shape of the potential (\ref{potxi}) is
completely determined. To compute the eigenvalues $\epsilon_\nu$ we
will truncate the infinite dimensional matrix of the operator
$\Delta^r_{\lambda \,k}$ to the $(N\times N)$ block. Thus, the
matrix elements, $\bra{n}{\cal X}\ket{m}$ can be derived one by one
using the relation (\ref{elem}) with the calculated generating
functional (\ref{ourgf}).

For each $k$, $N$ values of eigenvalues $\epsilon_n=2\nu_n +1$ will
be obtained; where $n=1,2,\ldots,N$. According to (\ref{nucond})
there are a finite number of values of $\nu_n$ which could determine
discrete energy levels, the other ones overlap the continuous
spectrum. Let us denote by $D$, the largest value for the discrete
part, $\nu_1,\nu_2,\ldots,\nu_D$ with $\epsilon_n\leq
-\lambda(k-1)$, equivalently we can write $E_n\in [0,\,M)$ for $n=\{
1,2,\ldots, D\}$. In what follows we will denote the higher energy
from the discrete spectra with $\varepsilon_d$.

Calculations have been made, using symbolic and numeric computation.
Thus we have analyzed the case with $N=50$, calculating the
dependence of $\nu_n$ versus $n$ for the following values of
$k=\{11,21,41,61\}$ and with different values of
$\lambda=\{-1,-7,-10\}$.

\begin{figure}[h]
\centerline{\psfig{file=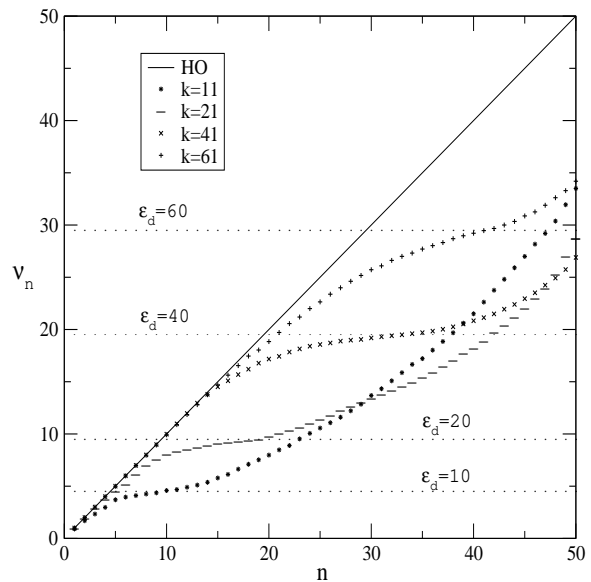,height=3in,width=3in}} \caption
{The dependence of pseuoquantum number $\nu_n$ on HO quantum number
$n$ with different values for $k=\{11,21,41,61,\}$ and $\lambda=-1$
in the order $r=3$ using a matix with $N=50$.}
 \label{depk}
\end{figure}

In the figure (\ref{depk}) we have represented the dependence of
pseuoquantum number $\nu_n$ versus HO quantum number $n$, in the
case of $\lambda =-1$. In this case the upper limit for the
eigenvalues is $\epsilon_n \leq k-1$ and represent the separation
limit between discreet and continuous spectra. A inflection point is
observed near this limit. This means that the distance between
energy levels is lowering and is minimum at the separation level.
The results are similar in this case of $\lambda =-1$, with those
obtained in \cite{RGaussBag} where $\epsilon_n \leq k$.

\begin{figure}[!h]
{\psfig{file=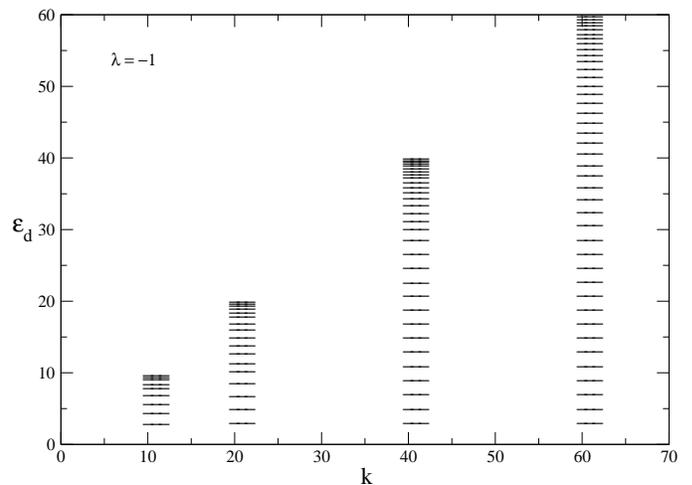,height=2.5in,width=3.5in}} \caption {The
dependence of eigenvalues $\epsilon_n$ with k in the order $r=3$
using a matix with $N=50$.}
 \label{evk}
\end{figure}

The eigenvalues ranging into the discrete domain $\epsilon_d=2\nu_d
+1$ with $d=\{1,2,\cdots, D\}$ are represented in figure
(\ref{evk}).

\begin{figure}[!h]
\centerline{\psfig{file=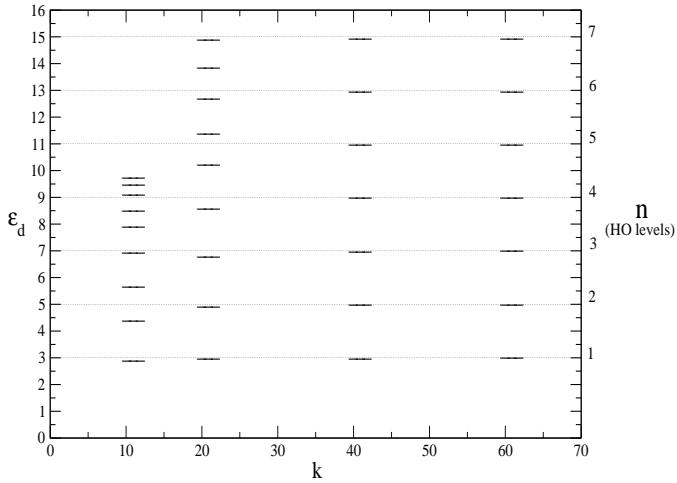,height=2.5in,width=3.5in}}
\caption {The comparison between calculated energy $\epsilon_d$ and
HO energy level $n=\{1,2,\cdots \}$}
 \label{evkD}
\end{figure}

The values of energy are under the values of HO, but are very close
to these, at least for the low energy levels. A closer look at the
disposal of the energies it is shown in the figure  (\ref{evkD}) the
amount of calculated energy are compared with those of HO.

\begin{figure}[h]
\centerline{\psfig{file=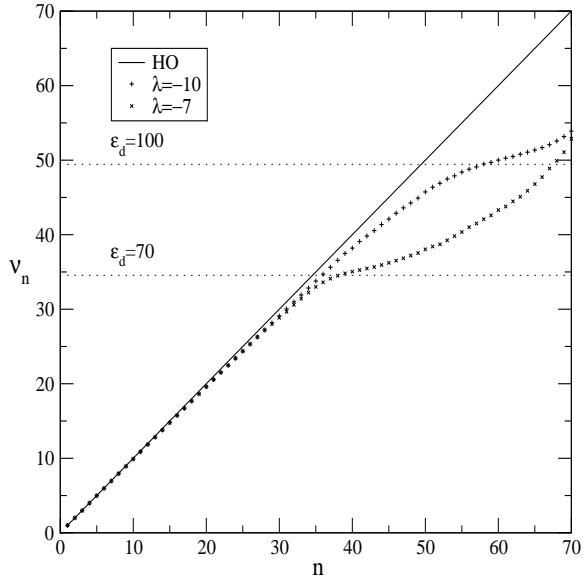,height=3in,width=3in}} \caption
{The dependence of pseuoquantum number $\nu_n$ on HO quantum number
$n$ with  $\lambda=\{-7,-10\}$ and $k=10$ in the order $r=3$ using a
matix with $N=70$.}
 \label{depl}
\end{figure}

As one can expect increasing the $\lambda$, the energy levels are
closer to HO ones, so the system can be assimilate with more
accuracy with an HO system. This seems to be natural because
increasing the absolute value of $\lambda$ the shape of potential
(\ref{potxi}) approaching to HO ones. The dependence of $\nu_n$
versus $n$ for  $\lambda=\{-7,-10\}$ is presented in figure
(\ref{depl}).

\begin{figure}[!h]
{\psfig{file=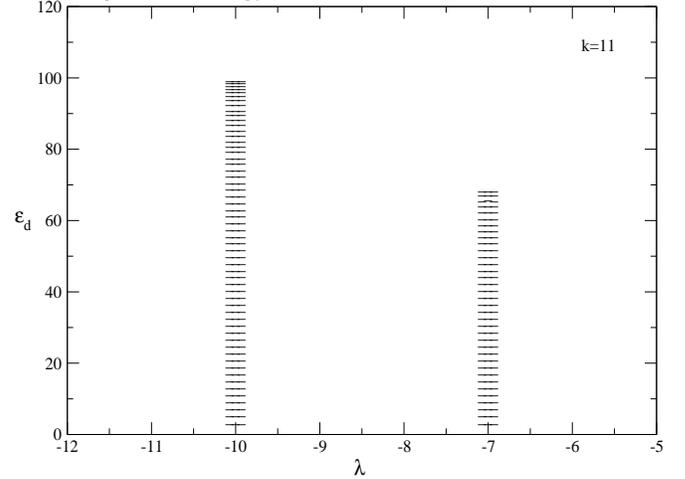,height=2.5in,width=3.4in}} \caption {The
dependence of eigenvalues $\epsilon_n$ with $\lambda$ in the order
$r=3$ using a matix with $N=70$.}
 \label{evl}
\end{figure}

The number of energy levels is increasing with $\lambda$, the
distance between levels remaining equidistant, less in the vicinity
of separation point. The eigenvalues ranging into the discrete
domain $\epsilon_d=2\nu_d +1$ with $d=\{1,2,\cdots, D\}$ are
represented in figure (\ref{evl}).

We have shown that a pseudo Gaussian bag with a massive quantum
particle inside, have an energy spectrum with a finite part $E_n\in
[0,\,M)$ and a continuous one $E_n\in [M,\,\infty)$. The energy
levels of the discrete spectra have been calculated in terms of
eigenvalues $\epsilon_d$, $d\in\{0,\ldots,D\}$ with the relation:
\begin{equation}
E_d=[M^2+M\omega(\epsilon_d+\lambda(\frac{M}{\omega}-1))]^\frac{1}{2}
\end{equation}
with a ground energy parameter $\lambda\leq-1$ and $M>\omega$. The
eigenvalues $\epsilon_n$ were calculated numerically, using the
method of generating functional.

The model is interesting from physical point of view, because it
offers the possibility to investigate small potential wells with
arbitrary energy levels from where the quantum particle can escape
when a transition from discrete to continuous spectrum becomes
possible. Starting with $\lambda=-1$ and $k=1$ where no energy
levels appears, any model $(\lambda,k)^r$ can be obtained with an
arbitrary energy levels. We think this model can be used to
calculate electronic transitions in semiconductors.

\subsection*{Acknowledgments}

We would like to thank  Ion Cot\u aescu for helpful comments and
suggestions that have improved the development of paper.


\begin{thebibliography}{0}

\bibitem{Yukawa} H.Yukawa, Phys. Rev. 91 (1953) 416.
\bibitem{oscil} Y.S.Kim and M.E.Noz, Am. J. Phys. 46 (1978) 480;
Y.S.Kim and E.P.Wigner, Phys. Rev. A38 (1988) 1159.
\bibitem{Mdo} M. Moshinsky and Szczepaniak 1989 J. Phys. A: Math. Gen. 22 L817-L819
\bibitem{NavarroA} V.Aldaya, J.Bisquert and J.Navarro-Salas, Phys. Lett. A156 (1991)
351. V.Aldaya, J.A. de Azc´arraga, J.Bisquert and J.M.Cerver´o,
J.Phys. A23 (1990) 707.
\bibitem{Nav} D.J.Navarro, J.Navarro-Salas, hep-th/9406001
\bibitem{CotRHO} I.I. Coatescu Int.J.Mod.Phys. A12 (1997) 3545-3550
\bibitem{CGaussBag} Ardelean, G.; Cotaescu, I. I.; Vulcanov, D. N. Phys.Lett.A, (2001)
283, 147-151.
\bibitem{RGaussBag} Ardelean, G.; Cotaescu I., Phys.Lett.A, (2003)
316, 168-172.
\bibitem{PGO}Cotaescu,I.; Gravila, P; Paulescu, M., Int.J.Mod.Phy. C, 19, 1607-1615 (2008).
\bibitem{Ball}L.B.Ballentine, Quantum Mechanics: A Modern Development
(WS. Publ. London, 2000).
\end{thebibliography}
\end{document}